\documentclass[aps,prb,reprint,twocolumn,superscriptaddress,showpacs,amssymb]{revtex4-1} 
\usepackage[english]{babel}
\usepackage{graphics}
\usepackage{graphicx}
\usepackage{epsfig}
\usepackage{amssymb}
\usepackage{dcolumn}
\usepackage{bm}
\usepackage{color}
\usepackage{lineno}
\usepackage{sidecap}
\usepackage{verbatim}  
\usepackage{vector}  
\usepackage{wrapfig}
\usepackage{enumerate}
\usepackage{sidecap}
\usepackage{amsmath}
\usepackage{hyperref}

\begin{document}

\title{MoTe$_2$: An uncompensated semimetal with extremely large magnetoresistance}

\author{S. Thirupathaiah}
\email{t.setti@sscu.iisc.ernet.in}
\affiliation{Solid State and Structural Chemistry Unit, Indian Institute of Science, Bangalore, Karnataka, 560012, India.}
\author{Rajveer Jha}
\affiliation{CCNH, Universidade Federal do ABC (UFABC), Santo Andre, SP, 09210-580 Brazil.}
\author{Banabir Pal}
\affiliation{Solid State and Structural Chemistry Unit, Indian Institute of Science, Bangalore, Karnataka, 560012, India.}
\author{J. S. Matias}
\affiliation{CCNH, Universidade Federal do ABC (UFABC), Santo Andre, SP, 09210-580 Brazil.}
\author{P. Kumar Das}
\affiliation{CNR-IOM, TASC Laboratory AREA Science Park-Basovizza, 34149 Trieste, Italy.}
\affiliation{International Centre for Theoretical Physics, Strada Costiera 11, 34100 Trieste, Italy.}
\author{P. K. Sivakumar}
\affiliation{Solid State and Structural Chemistry Unit, Indian Institute of Science, Bangalore, Karnataka, 560012, India.}
\author{I. Vobornik}
\affiliation{CNR-IOM, TASC Laboratory AREA Science Park-Basovizza, 34149 Trieste, Italy.}
\author{N. C. Plumb}
\affiliation{Paul Scherrer Institut, Swiss Light Source, CH-5232 Villigen PSI, Switzerland.}
\author{M. Shi}
\affiliation{Paul Scherrer Institut, Swiss Light Source, CH-5232 Villigen PSI, Switzerland.}
\author{R. A. Ribeiro }
\affiliation{CCNH, Universidade Federal do ABC (UFABC), Santo Andre, SP, 09210-580 Brazil.}
\author{D. D. Sarma}
\affiliation{Solid State and Structural Chemistry Unit, Indian Institute of Science, Bangalore, Karnataka, 560012, India.}

\date{\today}

\begin{abstract}
Transition-metal dichalcogenides (WTe$_2$ and MoTe$_2$) have drawn much attention, recently, because of the nonsaturating extremely large magnetoresistance (XMR) observed in these compounds in addition to the predictions of likely type-II Weyl semimetals.   Contrary to the topological insulators or Dirac semimetals where XMR is linearly dependent on the field, in WTe$_2$ and MoTe$_2$ the XMR is nonlinearly dependent on the field, suggesting an entirely different mechanism. Electron-hole compensation has been proposed as a mechanism of this nonsaturating XMR in WTe$_2$, while it is yet to be clear in the case of MoTe$_2$ which has an identical crystal structure of WTe$_2$ at low temperatures.   In this paper, we report low-energy electronic structure and Fermi surface topology of MoTe$_2$ using angle-resolved photoemission spectrometry (ARPES) technique and first-principle calculations, and compare them with that of WTe$_2$ to understand the mechanism of XMR. Our measurements demonstrate that MoTe$_2$ is an uncompensated semimetal,  contrary to WTe$_2$ in which compensated electron-hole pockets have been identified, ruling out the applicability of charge compensation theory for the nonsaturating XMR in MoTe$_2$. In this context, we also discuss the applicability of the existing other conjectures on the XMR of these compounds.
\end{abstract}
\pacs{}

\maketitle

Materials showing extremely large magnetoresistance (XMR) have potential applications in spintronics. Among them, the semimetals, WTe$_2$ and MoTe$_2$, have attracted  a great deal of research interests recently as they show nonsaturating extremely large MR~\cite{Ali2014, Zhou2016} even at 60 T of applied field in addition to the prediction of Weyl-nodes~\cite{Soluyanov2015, Sun2015a}. While a negative MR is widely known in many magnetic materials~\cite{Baibich1988, Binasch1989, Salamon2001}, positive MR has been noticed in some nonmagnetic materials~\cite{Schubnikow1930, Xu1997, Yang1999, Lee2002}. Some of these nonmagnetic compounds such as Ag$_{2+\delta}$Te/Se~\cite{Xu1997, Lee2002}, graphene~\cite{Singh2012},  Bi$_2$Te$_3$~\cite{Qu2010, Wang2012b}, and Cd$_3$As$_2$~\cite{Liang2014, Feng2015} show linearly field dependent MR, while type-II Weyl semimetals (WTe$_2$ and MoTe$_2$)~\cite{Ali2014, Zhou2016}, LaSb~\cite{Zeng2016} and ZrSiS~\cite{Lv2016} show quadratic dependence of MR on the field.

Charge compensation is explained as a mechanism of nonsaturating XMR in the compounds showing quadratic field dependent MR, while nontrivial band topology is thought to be responsible for the same in compounds showing linear field dependent MR. An ARPES report on WTe$_2$ demonstrated temperature dependent band structure that is consistent with the temperature dependent MR~\cite{Pletikosic2014}, thus supporting the conjecture of the charge compensation~\cite{Ali2014},  while the other ARPES reports point to the importance of the spin orbit coupling and the impact of the thickness dependence of the charge compensation~\cite{Jiang2015, Das2016}. An ARPES report on LaSb showed temperature independent band structure, while MR is still temperature dependent~\cite{Zeng2016}. Interestingly, a recent ARPES report on YSb has pointed that these two mechanisms could not explain the observed XMR in YSb which is neither a topological semimetal nor a Weyl semimetal~\cite{He2016}. All of these experimental observations are clearly demonstrating that there is no consensus yet on the mechanism of XMR in the nonmagnetic semimetals.

  In this paper, we report the electronic structure of MoTe$_2$ using high-resolution angle-resolved photoemission spectroscopy  and first-principle calculations. Though there exist few ARPES reports on MoTe$_2$~\cite{Jiang2017, Xu2016, Tamai2016, Deng2016, Huang2016, Liang2016}, discussing mostly the  predictions of Weyl nodes, no ARPES report has addressed till date on the origin of XMR in MoTe$_2$. Here, we show that although the low temperature crystal structure of both MoTe$_2$ and WTe$_2$ is identical, the bulk electronic structure of MoTe$_2$ is markedly different from WTe$_2$. We found three interconnected hole pockets and four disconnected electron pockets in MoTe$_2$ in our  ARPES studies which are qualitatively supported by our density functional theory (DFT) calculations. Previous ARPES reports on MoTe$_2$ showed electron pockets that are located only along the $\Gamma-X$ direction. In our ARPES measurements and DFT calculations we realized another pair of electron pockets located at each of the $Y$-points. We further noticed temperature independent band structure when measured between 20 and 130 K.  As the net size of hole pockets is larger than that of the net size of electron pockets in addition to the temperature independent band structure, the conjecture of charge compensation appears to be invalid for MoTe$_2$ in explaining the property of nonsaturating XMR.  Thus, our electronic structure studies of MoTe$_2$ rule out the mechanism of charge compensation as a cause of the unsaturated XMR in these compounds.

\begin{figure}[t]
	\centering
		\includegraphics[width=0.45\textwidth]{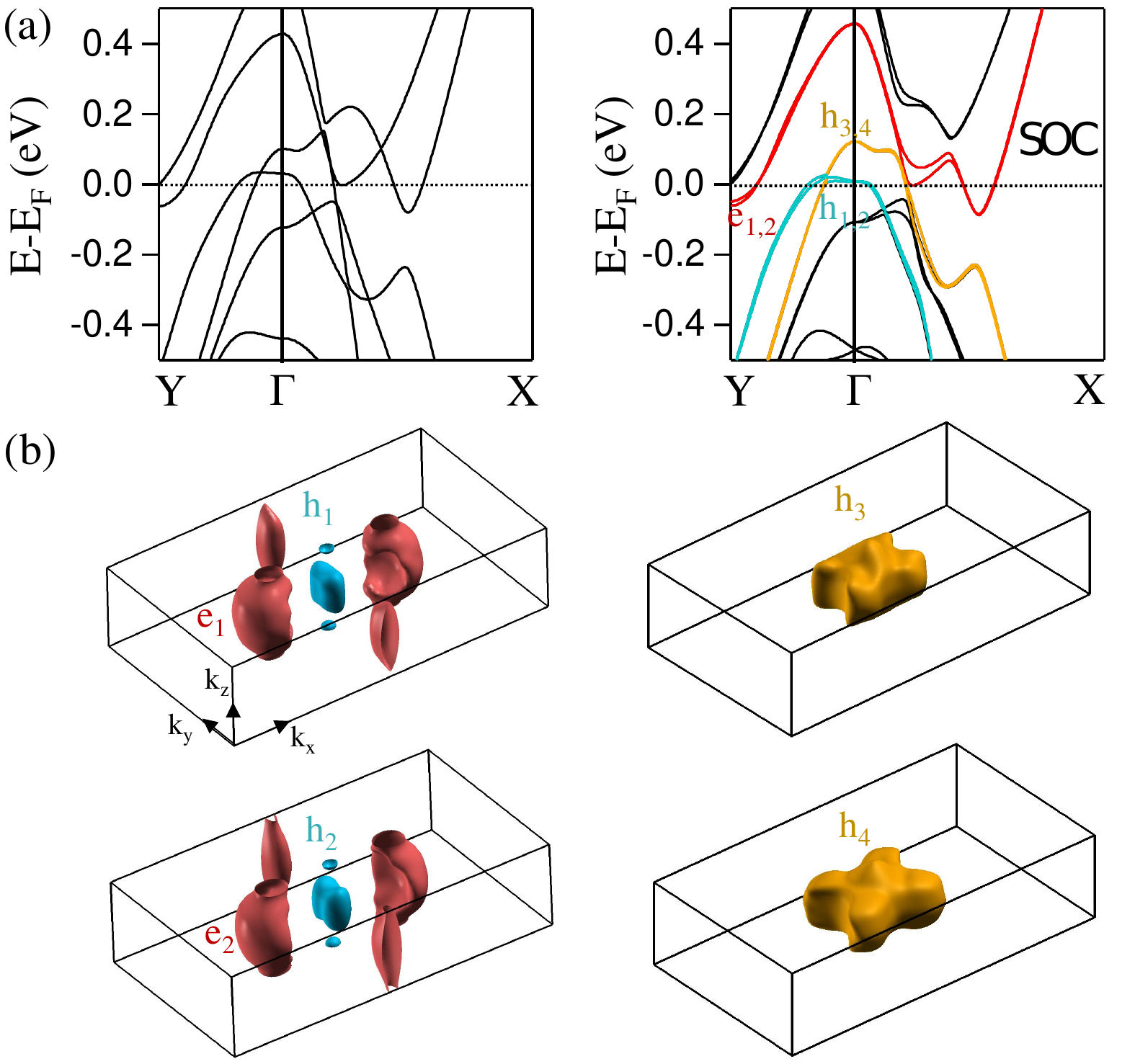}
	\caption{(a) Calculated band structure of MoTe$_2$ without (left panel) and with spin-orbit coupling (right panel). (b) 3D view of the calculated Fermi surfaces with spin-orbit coupling. See Ref.~\onlinecite{Calculations} for details on the first-principle calculations}
	\label{1}
\end{figure}

Good quality single crystals of stoichiometric MoTe$_2$ were grown using self-flux at Universidade Federal do ABC (UFABC), Brazil as discussed in Ref.~\onlinecite{Zhou2016}. The crystals have a platelet-like shape with shiny surface. The crystals were structurally characterized using powder X-ray diﬀraction to conﬁrm the bulk purity and the monoclinic crystal structure with P2$_1$/m1 space group [see Fig.S1 (b)]. ARPES measurements were performed in Swiss Light Source (SLS) at the SIS  beamline using a VG-Scienta R4000 electron analyzer and at the APE beamline in Elettra Synchrotron,  Trieste equipped with a Scienta DA30 deflection analyzer. The angular resolution was set at $0.2^\circ$ for R4000 and at $0.3^\circ$ for DA30. The used photon energy ranged between 20 and 45 eV. Overall energy resolution was set between 15 and 25~meV depending on the photon energy and beamline employed.  Samples were cleaved $\textit{in situ}$ at a temperature of 20 K and the chamber vacuum was better than $5\times10^{-11}$ mbar during the measurements. ARPES measurements were performed on two samples broken from a single big crystal, named here as A and B to differentiate from each.

\begin{figure*}
	\centering
		\includegraphics[width=0.95\textwidth]{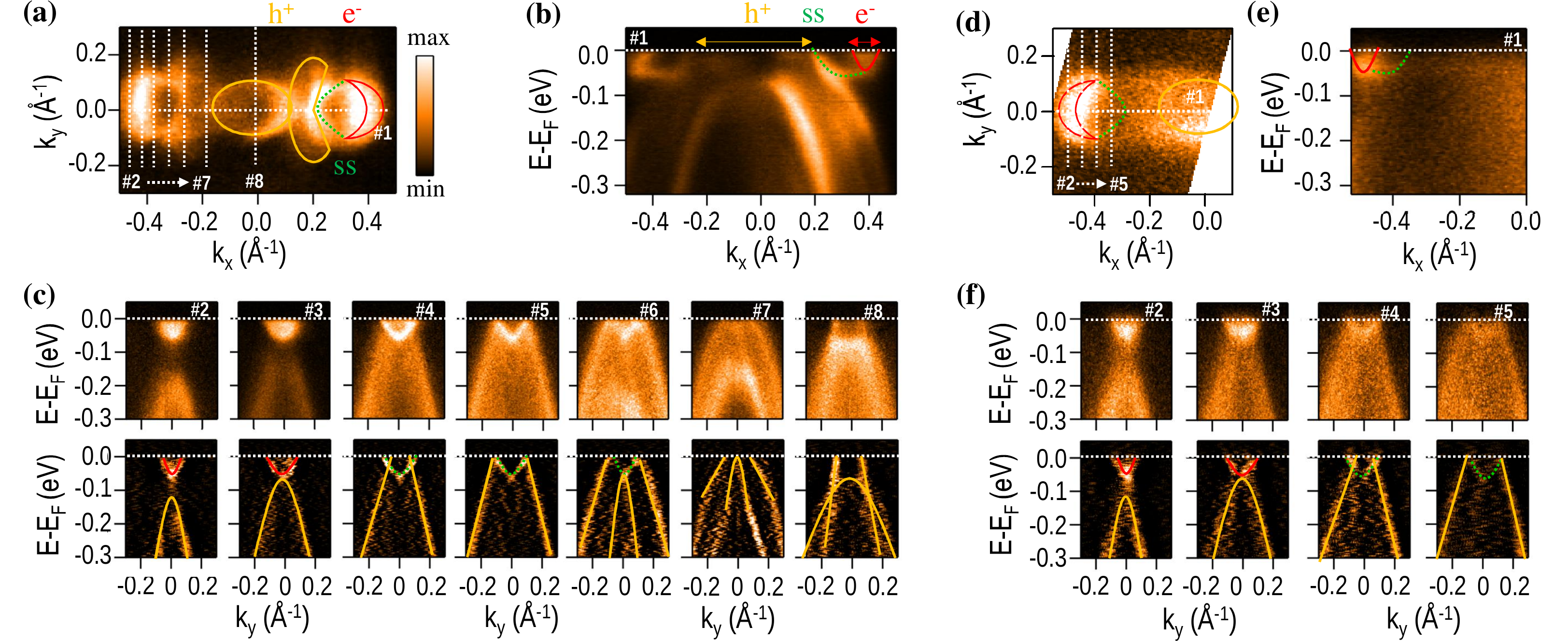}
	\caption{ ARPES data of MoTe$_2$ measured on sample A.  All the data are measured using $p$-polarized light with a photon energy of h$\nu$=20 eV. Note here that the 20 eV photon energy extracts the bands from $k_z$ = 0 plane~\cite{Xu2016, Tamai2016}. The data shown in (a)-(c) are measured at a sample temperature of 20 K. (a) depicts Fermi surface (FS) map.  Solid curves represent bulk Fermi sheets, while dashed curve represents Fermi arc from the surface. (b) shows energy distribution map (EDM) taken along the cut \#1 as shown on FS map. Top panels in (c) show EDMs taken along the cuts \#2 $-$ \#8 from left to right, respectively. Bottom panels in (c) are respective second derivatives of the EDMs shown in the top panels.  (d)-(f) depict similar data of (a)-(c) except that these are measured at 130 K. On the FS maps the hole (yellow) and electron pockets (red) contributed from bulk  are schematically shown by solid contours and green color dashed contours show the contribution from surface.}
	\label{2}
\end{figure*}

In the energy ($E$)-momentum ($k$) plot [see left panel in Figure~\ref{1}(a)] obtained from the DFT calculations without spin-orbit interaction we identify two holelike bands dispersing along the $\Gamma-X$ and $\Gamma-Y$ directions, while an  electronlike band dispersing along the $\Gamma-Y$ direction and another electronlike band dispersing along the $\Gamma-X$ direction. Interesting to see from Fig.~\ref{1}(a) that one of the hole bands shows almost flat dispersion along $\Gamma-X$ and $\Gamma-Y$ directions over a range of crystal momentum, whose band top is in vicinity of the Fermi level ($E_F$). From the $E-k$ plot obtained with spin-orbit interactions [see right panel in Figure~\ref{1}(a)], we identify that the holelike and electronlike bands are split resulting into two sets of hole and electron pockets with slightly different sizes as shown in Fig.~\ref{1}(b). There, we could find four electron pockets and four hole pockets. These calculations are in very good agreement with the calculations reported in Ref.~\onlinecite{Zhou2016, Wang2016}, but not consistent with the calculations reported in Ref.~\onlinecite{Qi2016}. We further noticed that the number of electron pockets is sensitive to the position of $E_F$. A small increment of the Fermi level leads to the presence of several tiny electron pockets as reported in Ref.~\onlinecite{Zhou2016}.

 In Figure~\ref{2} we show ARPES data of MoTe$_2$ measured on the sample A. In the Fermi surface map shown in Fig.~\ref{2}(a),  we can identify three well-connected, two jelly bean-shaped and one oval-shaped, hole pockets around the zone center ($\Gamma$) and two crescent-shaped electron pockets located along the $k_x$ direction. It is also clear from the same map that the Fermi topology of these compounds is highly anisotropic, that means,  spectral intensity distribution along the $k_x$ direction is entirely different from that along the $k_y$ direction. This observation is in-line with the anisotropy of crystal structure as well [see Figure S1 in the supplemental material]. We further noticed an electronlike Fermi arc connecting both the bulk hole and electron pockets as shown by green dashed curve in Fig.~\ref{2}(a). This Fermi arc is ascribed to the presence of Weyl nodes in MoTe$_2$~\cite{Sun2015a, Wang2016}.  The three hole pockets are related to the $h_3$ ( $h_4$) Fermi sheet as shown in Fig.~\ref{1}(b). We could not resolve the $h_1$ ($h_2$) hole pocket due to the matrix element effects.

To elucidate further the nature of electronlike and holelike band dispersions,  we made cuts along the $k_x$ and $k_y$ directions as shown in Fig.~\ref{2} (a). From Fig.~\ref{2}(b), EDM cut taken in the $k_x$ direciton,  one can notice that the electronlike surface state disperses in such way that connecting the bottom of the bulk electronlike band and the top of bulk holelike band. As the band structure of these compounds is complex near the Fermi level it is difficult to disentangle the individual bands. From the EDM cuts taken in the $k_y$ direction,  we identified bulk electronlike band dispersion [see cuts~\#2 and~\#3 in Fig.~\ref{2}(c)] and surface electronlike band dispersion [see cuts~\#4 $-$ \#6 in Fig.~\ref{2}(c)]. Similarly, holelike bands are seen from the cuts~\#7 and \#8. These observations are consistent with the existing ARPES reports on MoTe$_2$~\cite{Jiang2017, Xu2016, Tamai2016, Deng2016, Huang2016, Liang2016}. Figures~\ref{2} (d)-(f) depict similar data shown in  Figs.~\ref{2} (a)-(c) but measured at a sample temperature of 130 K. From  Figs.~\ref{2} (d)-(f) it is clear that the band structure of MoTe$_2$ is largely intact with increase in temperature except that the bands are diffused due to thermal broadening. Nevertheless due to this thermal broadening, we were unable to find the jelly bean-shaped hole pockets from Fig.~\ref{2}(d) which has relatively reduced spectral intensity than the other Fermi sheets [see Fig.~\ref{2}(a)]. Nevertheless, the electronlike surface state is visible even at 130 K.

\begin{figure} [ht]
	\centering
		\includegraphics[width=0.45\textwidth]{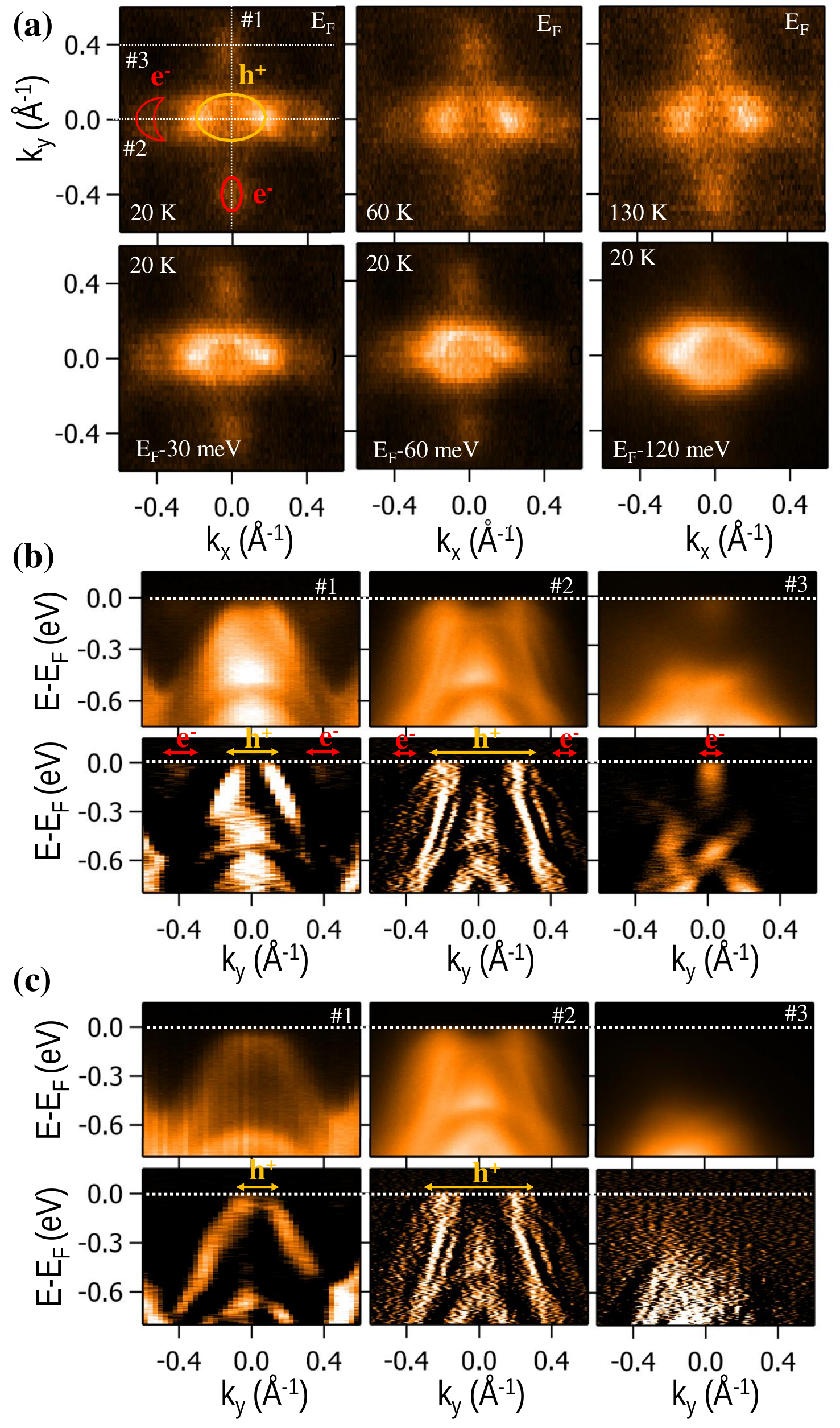}
	\caption{ARPES data of MoTe$_2$ measured on sample B. All the data are measured with a photon energy of 45 eV. Top panels in (a) depict FS maps measured at sample temperature of 20, 60, and 130 K. Bottom panels in (a) depict constant energy contours taken at a binding energy E$_B$ = 30, 60, and 120 meV at 20 K. In top panels of (b) we show EDMs taken along the cuts \#1 $-$ \#3 as shown on the FS map. Bottom panels of (b) are the second derivatives of respective EDMs shown in the top panels. (c) depicts similar data of (b) but measured with $s$-polarized light,  while the data shown in (a) and (b) are measured with $p$-polarized light. Please see Figure S2 given in the supplementary information for the definition of $p$-and $s$-polarized lights.}
	\label{3}
\end{figure}

 In Figure~\ref{3} we show the ARPES data of MoTe$_2$ measured on sample B. Using inner potential of 13 eV, we calculated that 45 eV photon energy extracts the bands from $k_z$ = 0.1 ($\pi/c$) plane.  From the maps shown in top panels of Fig.~\ref{3}(a), we could identify one holelike Fermi sheet (oval-shaped) at the zone center and two electronlike Fermi sheets (crescent-shaped) on either side of the hole pocket located along the $k_x$ direction. Surprisingly, we noticed two more oval-shaped Fermi sheets on either side of the hole pocket located at the $Y$-points dispersing in the $k_y$ direction. This Fermi sheet is not experimentally reported so far either from  ARPES measurements~\cite{Jiang2017, Xu2016, Tamai2016, Deng2016, Huang2016, Liang2016} or from Quantum Oscillations~\cite{Zhou2016} on MoTe$_2$,  though is found in our DFT calculations[see Fig.~\ref{1}(a)] and as well in the reported DFT calculations~\cite{Zhou2016, Wang2016}.  To further elucidate nature of these Fermi sheets, we made EDM cuts in the $k_x$ and $k_y$ directions as shown in the map [see Fig.~\ref{3}(a)]. From cut~\#3 it is clear that the Fermi sheet shows electronlike band dispersion with a band bottom at $\approx$ 120 meV below $E_F$. Same is confirmed from the constant energy contours [see bottom panels in Fig.~\ref{3}(a)] that the intensity of electron pockets at $Y$-point disappear in the contour taken at a binding energy of 120 meV below $E_F$. We further examined that this electron pocket is absent when measured with $s$-polarized light [see Fig.~\ref{3}(c)].  From the cuts~\#1 and \#2 we again found holelike band dispersions at the $\Gamma$-point using both $p$-and $s$-polarized lights,  consistent with the data of the sample A. However, to emphasize, we found only one oval-shaped hole pocket from the sample B, contrary to the sample A in which we detected three hole pockets (two jelly bean-shaped and one oval-shaped). Moreover, we did not find the electronlike surface state from the sample B.  All this simply demonstrate that the cleavage plane of sample A is different from that of sample B (see also Figure S2 in supplemental material). This finding is consistent with the photoemission observations made on WTe$_2$~\cite{Wu2016, Bruno2016} where it was showed that the differing cleavage planes lead to the differing band structures.  Temperature dependent data of the Sample B is again consistent with sample A, i.e., the band structure of sample B is mostly intact with temperature when measured between 20 and 130 K. As can be seen from the top panels of Fig.~\ref{3} (a), on increasing the sample temperature from 20 K to 130 K, apart from trivial thermal broadening of the Fermi sheets we hardly detect any changes in the band structure. Further details on the orbital charter of these bands can be found in Figure S3 (supplemental).

 From the data taken on sample A and B, we found  a momentum vector of 0.09 $\AA^{-1}$ for the crescent-shaped electron pocket in the $k_y$ direction,  momentum vectors of 0.13 $\AA^{-1}$ and 0.07 $\AA^{-1}$ for the oval-shaped hole pocket in the $k_x$ and $k_y$ directions, respectively and momentum vectors of 0.13 $\AA^{-1}$ and 0.04 $\AA^{-1}$ for the jelly-bean shaped hole pocket in the $k_x$ and $k_y$ directions, respectively. With the help of these momentum vectors and following the method shown in supplemental materials,  we have calculated the number of electron ($n_e$) and hole ($n_h$) carriers, equivalent to 0.014 and 0.026 per unit cell, respectively. These values are estimated qualitatively without taking the $k_z$ dependence of band structure and considering that the crescent-shaped electron and jelly bean-shaped hole pockets as half-cylinder and half-elliptical cylinder, respectively and the oval-shaped hole pocket is considered as an elliptical cylinder in three dimensional.  Our estimate of $n_e$ = 0.014/unit cell (4 $\times$ 10$^{19}$ cm$^{-3}$) is in good agreement with the electron density values of 0.011/unit cell (2 $\times$ 10$^{19}$ cm$^{-3}$) ~\cite{Zhou2016} and 0.0153/unit cell (5 $\times$ 10$^{19}$ cm$^{-3}$)~\cite{Qi2016} reported on MoTe$_2$ at low temperatures using the quantum Hall measurements.

As reported here,  the electronic structure of MoTe$_2$ is temperature independent in addition to the charge carrier imbalance. Therefore, the proposed theory of charge compensation for the nonsaturating XMR in these compounds seems to be invalid. A similar conclusion is arrived in Ref.~\onlinecite{Zandt2007} using Hall probe, but lacked a quantitative analysis as we did here.  In addition to the charge compensation the other mechanisms are,  (i) magnetic field induced changes of the band structure~\cite{Wang2014}, (ii) spin-orbit interaction~\cite{Jiang2015, Das2016}, and (iii) nontrivial band topology with the time reversal symmetry breaking~\cite{Tafti2015}. Point (iii) is not applicable to these semimetals as these (MoTe$_2$) show trivial band topology~\cite{Tamai2016}. In the absence of magnetic field, it is highly unlikely that the spin-orbit interaction changes with temperature unless there is a structural transition which occurs only at 260 K. As can be seen in Fig.~\ref{1}(a), one of the two holelike bands is flat up to a range of crystal momentum around the $\Gamma$-point in the $k_x$ and $k_y$ directions. Thus, hinting at more localized hole pockets when compared to the electron pockets which are very dispersive near the Fermi level. This is in agreement with previous reports which showed that the electron carrier mobility ($\mu_e$) is two times higher than the hole carrier mobility ($\mu_h$) in MoTe$_2$~\cite{Zhou2016}. Hence, these compounds  have dominant electron-type transport in MoTe$_2$~\cite{Keum2015, Qi2016, Zhou2016} although, as we showed  above, the hole concentration is higher than the electron concentration. After ruling out point (iii) and charger carrier compensation as the mechanisms of nonsaturating XMR in MoTe$_2$, we suggest that the combination of Fermi surface deformation and spin-orbit interactions in the presence of external magnetic field could be a plausible mechanism. We also do not rule out a significant role played by the differing mobilities of holes and electrons~\cite{He2016}.


In conclusion, we studied the low-energy band structure of MoTe$_2$ semimetal by means of ARPES technique and DFT calculations.  The ARPES data on two different samples from the same preparation batch demonstrate that the band structure is cleavage dependent, similar to what was recently observed in WTe$_2$~\cite{Wu2016}. Overall, the experimental findings are quantitatively consistent with our DFT calculations. Our results further demonstrate that MoTe$_2$ is a non-compensated semimetal as we found qualitatively that the number of hole carriers ($n_h$) is higher than electron carriers ($n_e$). This observation invalidates the theory of charge compensation for the nonsaturating XMR in MoTe$_2$. Temperature independent band structure of MoTe$_2$ is adding further difficulties in understanding the temperature dependent XMR recorded in MoTe$_2$. We believe, our results of MoTe$_2$ present an invaluable information to the emerging field of XMR physics in type-II Weyl semimetals and suggest a revision on the current understanding of the nonsatuarting XMR of these compounds.


S.T.  acknowledges support by the Department of Science and Technology, India through the INSPIRE-Faculty program (Grant number: IFA14 PH-86). The authors thank the Department of Science and Technology, India (SR/NM/Z-07/2015) for the financial support and Jawaharlal Nehru Centre for Advanced Scientific Research (JNCASR) for managing the project. The authors acknowledge the financial support given for the measurements at Elettra Synchrotron under Indo-Italian (DST-ICTP) cooperation Program. R.A.R. acknowledges support by FAPESP (grant no. 2011/19924-2). This work has been partly performed in the framework of the nanoscience foundry and fine analysis (NFFA-MIUR Italy) project.

\bibliography{MoTe2}

\section{Carrier density calculations}
According to the Luttinger's theorem~\cite{Luttinger1960},  the carrier density is directly proportional to the volume enclosed by the Fermi surface. Therefore, the equation for carrier density can be written in a simplified form,
\begin{equation}
n_{e,h}= \frac{V_{FS}}{V_{BZ}},
\end{equation}
where $V_{FS}$= volume of the Fermi surface and $V_{BZ}$=$\frac{(2\pi)^3}{a b c}$, volume of the Brillouin zone (BZ). Here $a$, $b$ and $c$ are the lattice constants.

In the present case we considered crescent-shaped electron pocket as half-cylinder in three dimensional (3D) in the case of $k_z$ independent band structure. Then, the volume is given by
\begin{equation}
V_{FS}^{e} = \frac{\pi r^2 h} {2} = \frac{(k_F \pi)^2} {2c}
\end{equation}
Here $k_F$ is the momentum vector. There are four of such electron pockets per Brillouin zone (see Fig. 1 (b) in the main paper) after considering the spin-orbit coupling (SOC).

Similarly, we considered jelly-bean shaped hole pocket as half-elliptical cylinder, and the volume is given by
\begin{equation}
V_{FS}^{h1} = \frac{\pi r_1 r_2 h} {2} = \frac{k_{F}^{x} k_{F}^{y} (\pi)^2} {2c}
\end{equation}
There are four of these hole pockets per Brillouin zone after SOC.

We considered oval shaped hole pocket as an elliptical cylinder, and the volume is given by
\begin{equation}
V_{FS}^{h2} = \pi r_1 r_2 h =   \frac{k_{F}^{x} k_{F}^{y} (\pi)^2} {2c}
\end{equation}
There are two of these hole pockets per Brillouin zone after SOC.

Finally,  the total electron density is given by
\begin{equation}
n_{e}= 4 \frac{V_{FS}^{e}}{V_{BZ}} = 4 \frac{(k_F \pi)^2} {2c} \frac{a b c}{(2\pi)^3}= \frac{ab(k_F)^2} {4\pi}=0.014/unit~ cell
\end{equation}
with $k_F$=0.09$\AA^{-1}$, $a$= 3.477 $\AA$ and $b$=6.355 $\AA$

and the hole density is given by
\begin{equation}
n_{h}= \frac{4 V_{FS}^{h1} +2 V_{FS}^{h2}}{V_{BZ}}=0.026/unit~cell
\end{equation}
with $k_F^x$ = 0.13 $\AA^{-1}$ and $k_F^y$ = 0.09 $\AA^{-1}$ for the jelly-bean shaped hole pocket and $k_F^x$ = 0.13 $\AA^{-1}$ and $k_F^y$ = 0.07 $\AA^{-1}$ for the oval shaped hole pocket (see the main paper for more details).

\begin{figure*}
	\centering
		\includegraphics[width=0.90\textwidth]{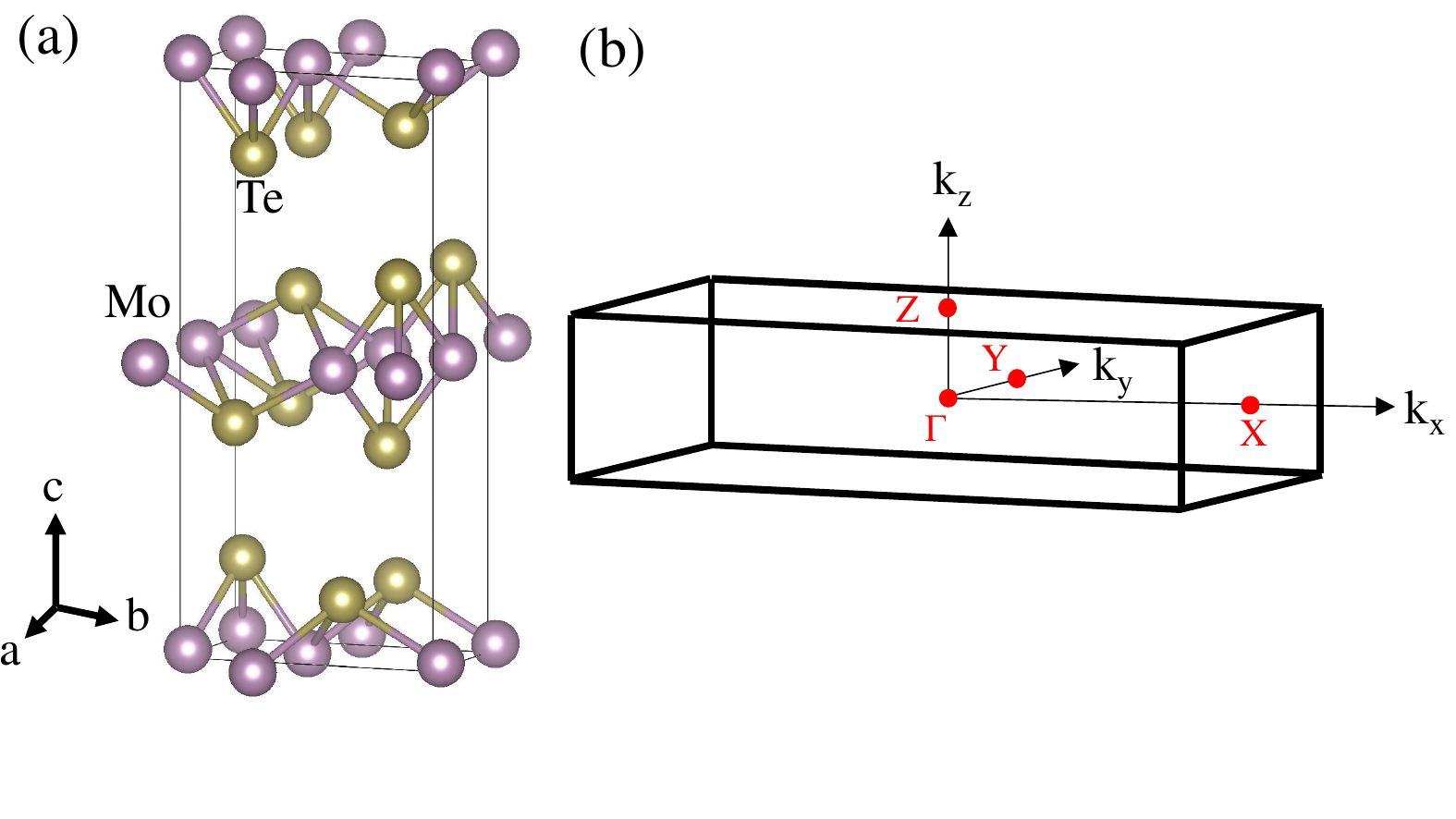}
	\caption{(a) Crystal structure of MoTe$_2$ and (b) 3D view of the Brillouin zone. (c) Powder X-ray diffraction pattern of cleaved MoTe$_2$ single crystal, high intensity peaks are well indexed with monoclinic crystal structure of space group P2$_1$/m1. (d) Temperature dependent electrical resistivity of MoTe$_2$ single crystals, current is applied along the $a$-axis. Hysteresis between warming and cooling cycle at T = 260K suggests a first order phase transition from $\beta$-MoTe$_2$ to T$_d$-MoTe$_2$.}
	\label{1}
\end{figure*}

\begin{figure*}
	\centering
		\includegraphics[width=0.99\textwidth]{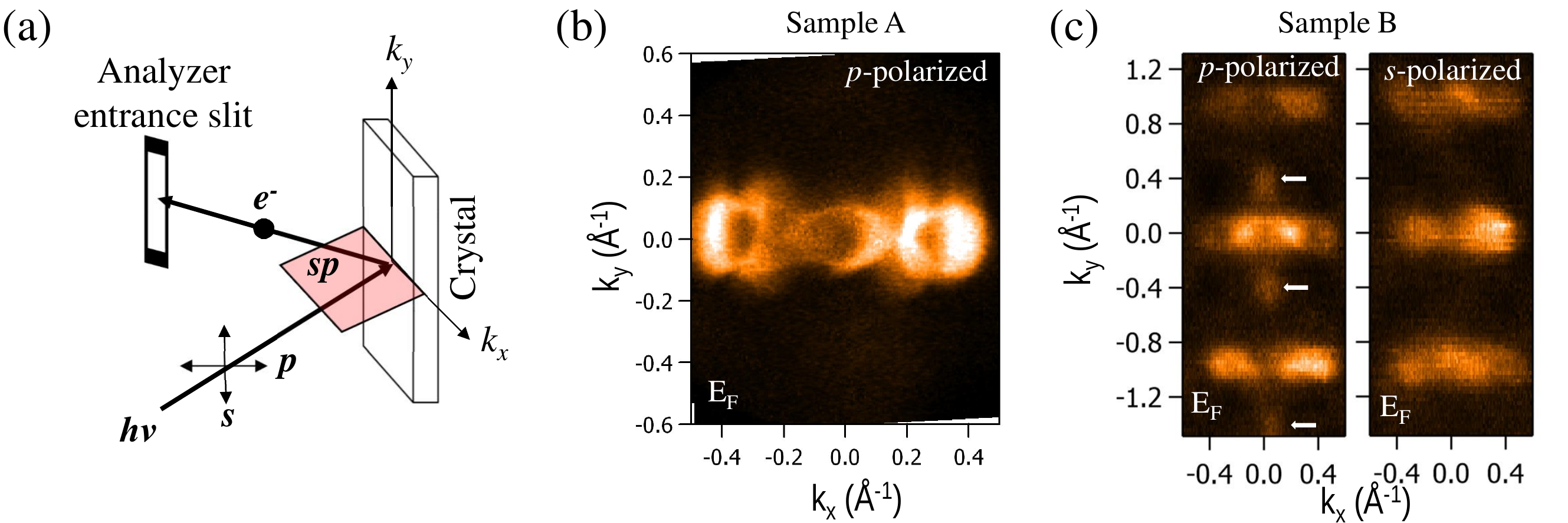}
	\caption{(a) Schematically shows a typical measuring geometry in which  $s$- and $p$-plane polarized lights are defined with respect to the analyzer entrance slit and the scattering plane (SP). (b) Shows Fermi surface map measured on sample A with a photon energy of 20 eV using $p$-polarized light. (c) shows Fermi surface maps measured on sample B with a photon energy of 45 eV using $p$-and $s$-polarized lights. Here one can clearly notice that the Fermi sheet shown by white arrows in the right panel of Fig.~1(c) is absent in sample A. Moreover, it also disappears when measured with $s$-polarized light on sample B.}
	\label{1}
\end{figure*}

\begin{figure*}
	\centering
		\includegraphics[width=0.99\textwidth]{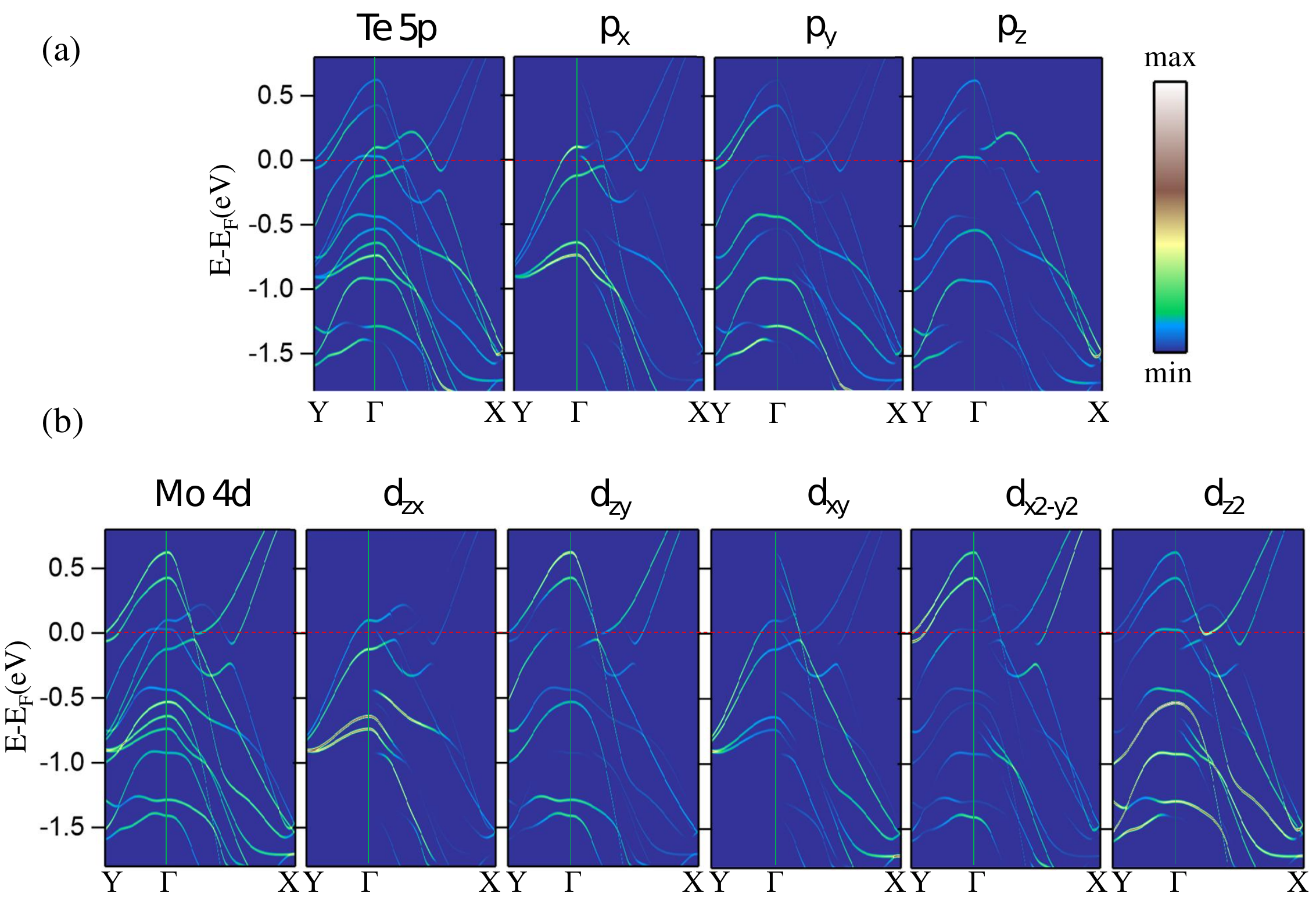}
	\caption{ Calculated band structure of MoTe$_2$ for the orbital characters Te-5$p$ (a) and Mo-4$d$ (b). As can be seen from the figure that the electron and hole pockets of MoTe$_2$ are resulted from a strong hybridization between Te-5$p$ and Mo-4$d$.}
	\label{1}
\end{figure*}

\end{document}